# Microstructural characterization of a Canadian oil sand

Doan[1,3] D.H., Delage[2] P., Nauroy[1] J.F., Tang[1] A.M., Youssef[2] S.

[1] IFP Energies nouvelles, 1-4 Av. du Bois Préau, F 92852 Rueil-Malmaison Cedex, France
j-francois.nauroy@ifpen.fr
[2] Ecole des Ponts ParisTech, Navier/ CERMES
6 et 8, av. Blaise Pascal, F 77455 Marne La Vallée cedex 2, France
delage@cermes.enpc.fr
[3] now in Fugro France
doan_hong_47xf@yahoo.fr

**Abstract**: The microstructure of oil sand samples extracted at a depth of 75 m from the estuarine Middle McMurray formation (Alberta, Canada) has been investigated by using high resolution 3D X-Ray microtomography (µCT) and Cryo Scanning Electron Microscopy (CryoSEM). µCT images evidenced some dense areas composed of highly angular grains surrounded by fluids that are separated by larger pores full of gas. 3D Image analysis provided in dense areas porosity values compatible with in-situ log data and macroscopic laboratory determinations, showing that they are representative of intact states. µCT hence provided some information on the morphology of the cracks and disturbance created by gas expansion. The CryoSEM technique, in which the sample is freeze fractured within the SEM chamber prior to observation, provided pictures in which the (frozen) bitumen clearly appears between the sand grains. No evidence of the existence of a thin connate water layer between grains and the bitumen, frequently mentioned in the literature, has been obtained. Bitumen appears to strongly adhere to the grains, with some grains completely being coated. The curved shape of some bitumen menisci suggests a bitumen wet behaviour.

*Key words*: oil sand, microstructure, X-Ray microtomography, Cryo Scanning Electron Microscopy, core disturbance, bitumen.

## Introduction

Unconventional hydrocarbon resources (heavy oil, extra heavy oil and bitumen) are mainly trapped in sand and sandstone reservoirs in Western Canada and Eastern Venezuela basins. The laboratory characterisation of oil sands is difficult because of the significant disturbance of core samples due to pressure release and gas expansion during extraction (e.g. Dusseault 1980). This disturbance drastically affects the petrophysical and geomechanical characteristics of core samples. The microstructure of oil sands from Athabasca (Alberta, Canada) has been considered by various authors (e.g. Dusseault and Morgenstern 1978; Takamura 1982; Czarnecki et al. 2005) who proposed a model describing the mutual organisation of sand grains, bitumen and water. However, it appeared that an important hypothesis of this model, i.e. the presence of a 10-15 nm thick layer of connate water along the grains surface with no direct contact between sand and bitumen, needed further experimental confirmation.

In this paper, the microstructure of core specimens from Athabasca was investigated by using high resolution 3D X-Ray microtomography (µCT) coupled with image analysis and by using a Cryo Scanning Electron Microscopy. These two advanced methods were used to better understand the natural microstructure of oil sand and the effects of gas expansion.

## Background

The world's largest oil sand deposits (Athabasca, Cold Lake and Peace River) are located in Alberta (Canada). In the Athabasca deposit, bitumen is hosted in the unconsolidated quartz sands of the 32 m thick McMurray Formation at depths ranging from 0 to 700 m (Butler 1997). This formation is underlain by shales and limestones of the Devonian carbonates Formation and overlaid by a caprock made up of the shales and sandstones of the Clearwater Formation (NEB 2000; Hein and Marsh 2008). The McMurray Formation is a fluvial-estuarine complex deposited 120 million years ago within an Early Cretaceous paleo-valley system that is characterised by significant heterogeneity (Butler 1997). The formation is commonly divided into three stratigraphic layers known as Lower, Middle, and Upper McMurray. The Lower McMurray Formation corresponds to high energy, sand dominated fluvial sediments whereas the Upper McMurray Formation corresponds to low energy, mud dominated and estuarine to shallow marine sediments. The Middle McMurray Formation contains the main bitumen bearing reservoir while the Lower part is generally water bearing. The Upper part typically contains gas with some residual bitumen saturations. Oils migrated from source rocks that were deepened near the Rocky Mountains during their formation. From that time, a variety of biological and physico-chemical processes (e.g. aerobic biodegradation, anaerobic alteration, hydro-dynamism) degraded the oils, removing the lighter portions and only leaving the heavier organic residues (Mathieu 2008). Canadian heavy oils and bitumens are characterized by low API gravity (< 10 °API) and very high viscosity (up to several millions cPo) at reservoir conditions under temperatures between 8°C and 20°C according to depth. The average bitumen composition is defined by 83.2 % carbon, 10.4 % hydrogen, 0.94 % oxygen, 0.36 % nitrogen and 4.8 % sulphur, along with some trace amounts of heavy metals such as vanadium, nickel and iron (NEB 2000).

Oil sands generally display a dense structure with many interpenetrative contacts and rough grain surfaces resulting from deep burial under temperatures higher than presently over long geological periods (120 millions years, see Dusseault and Morgenstern 1978; Dusseault 1980, Wong 2000). The status of water in oil sands has been discussed by many authors (including Dusseault and Morgenstern 1978; Takamura 1982; Butler 1997; Shuhua and Jialin 1997; Czarnecki et al. 2005) who mentioned that water in oil sands could be present under three forms: at the grain-to-grain contacts, within the fines clusters and around the sand grains under the form of a thin water layer (about 10 nm thick). The hypothesis of a thin water layer around the grains, that corresponds to water wet behaviour, could explain the noticeable easiness of oil removal by using hot water processes that is typical of the McMurray formation. Czarnecki et al. (2005) underlined however that there is no clear experimental evidence about this "reasonable postulate" and that further investigation was required for confirmation. They also mentioned that Zajic et al. (1981) examined freeze fractured samples by transmission electron microscopy (TEM) without finding any evidence of such a water sheath around the grains. Conversely, an oil-wet behaviour could be due to organic coatings along the grain surfaces by humic matter or asphaltenic type molecules (Czarnecki et al. 2005). Based on Environmental Scanning Electron Microscopy observation, Schmitt (2005) suggested that the heavy oil components could act as a cementing amorphous agent binding the sand grains together. In their model of the microstructure of oil sand, Dusseault and Morgenstern (1978) also mention the presence of gas bubbles occluded in the bitumen.

As commented in detail by Dusseault (1980), a specificity of oil-sands is that their investigation in the laboratory is made quite delicate due to the significant disturbance that affects core samples. Disturbance is mainly due to the significant expansion (possibly higher than 15 %) of the considerable amount of gas dissolved in the fluid phase of oil sands (mainly methane, carbone dioxide and nitrogen) whereas oil sand do not contain any gas under in-situ stress and pore pressure conditions (Dusseault 1980; Dusseault and van Domselaar, 1982). Accounting for disturbance effects, Dusseault (1980) provided the following estimated in-situ values of the physical properties of an oil-rich Athabasca oil sand specimen cored at 40 m: a porosity of the order of 30 % with a fluid saturation of about 99-100 %, an oil content (% of total weight) of 14.3 % and a water content of 1.6 %. The corresponding values obtained from a "relatively high quality" specimen are in the range 30-33 % for porosity, 80-98 % for fluid saturation, 14 % oil content and 2-3 % water content. Due to the effects of gas expansion, Dusseault (1980) also estimated a dramatic decrease of one order of magnitude in Young's modulus (from 1400 to 140 MPa) and water permeability (between $10^{-10}$ - $10^{-12}$ m/s to $10^{-9}$ - $10^{-11}$ m/s).

Given that in-situ geophysical logs provide acceptable in-situ values of the physical properties of oil sands, Dusseault (1980) and Dusseault and van Domselaar (1982) proposed to quantify the quality of cored specimen by considering a disturbance index $I_D$ corresponding to the relative difference between the in-situ porosity $\phi$ and the porosity measured in the laboratory $\phi_e$ as follows:

[1] $$I_D = \left(\frac{\phi_e - \phi}{\phi}\right)100\%$$

The classification of specimen according to the value of index $I_D$ is given in **Erreur ! Source du renvoi introuvable.**.

Table 1. Assessment of oil sand specimens quality (after Dusseault and van Domselaar 1982)

| $I_D$ (%) | Specimen quality | Use |
|---|---|---|
| $I_D$ < 10 | Intact or slightly disturbed | High quality geomechanical tests |
| 10 < $I_D$ < 20 | Intermediate disturbance | Most petrophysical research |
| 20 < $I_D$ < 40 | Highly disturbed | Most petrophysical research |
| 40 < $I_D$ | Disrupted generally | - |

## Basic physical properties

The samples used in this work come from 80 mm diameter cores of 1.5 m length that have been extracted from the estuarine Middle McMurray formation at a depth of 75 m. Petrophysical log results carried out in the borehole from which the sample has been cored indicated an in-situ porosity in the range 31 - 35 % with in-situ permeabilities between 3 and 4 D. At 75 m depth, the vertical stress calculated from log data is equal to 1 694 kPa with a pore fluid pressure of 370 kPa (water table at 40 m below surface) resulting in a vertical in-situ effective stress of 1 324 kPa.

The cores were stored in the laboratory at a temperature of 10°C. The mineralogical composition was analysed on a washed specimen by using X-Ray diffraction (XRD). Washing was performed in a Soxlhet cell at 250°C with a solution consisting of 70 % toluene and 30 % methanol. XRD analysis showed that the sample was composed of close to 99 % of quartz. The grain size distribution of the sand (Figure 1) was determined by sieving and by using a laser granulometer. Both methods confirmed that the sand is moderately to well sorted, with a mean grain size ranging from 0.160 to 0.200 mm.

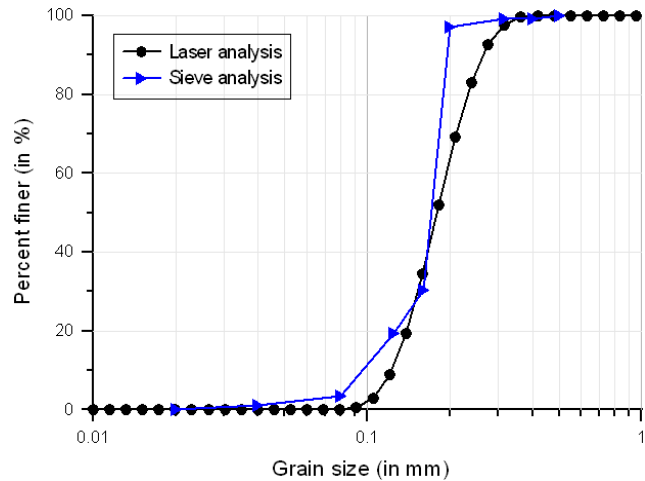

**Figure 1. Grain size distribution curves of the Athabasca oil sand**

Cylindrical specimens were trimmed by using a metal cylinder with a cutting edge (internal diameter 38 mm). The upper and lower faces were sawed, the sample was frozen and carefully polished on both faces by using sand paper. Samples were passed through a medical X-Ray scanner to investigate any possible disturbance. As shown in Figure 2, samples exhibited significant disturbance that appeared to be either diffuse (Figure 2a) or localised and crack-shaped (Figure 2b). As quoted by Dusseault (1980), gas expansion is a predominant source of disturbance, together with stress release, storage, transport and trimming. It is obvious that the 0.8 - 1.3 mm thick cracks of Figure 2b would have a dramatic impact on the geomechanical characteristics of the specimen with also a calculated porosity much higher than the natural one.

Trimmed specimens were frozen at -25°C in a domestic freezer prior to testing.

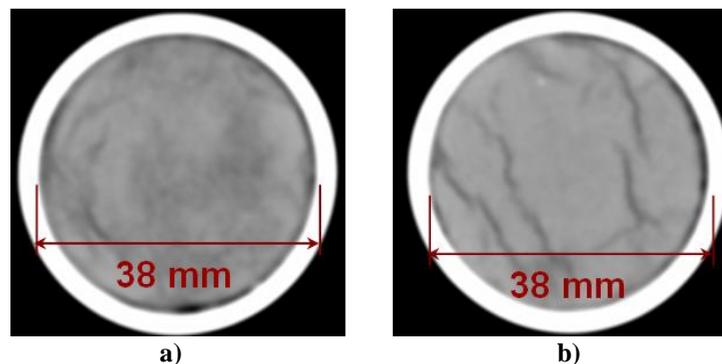

a)    b)

**Figure 2. Typical X-ray CT images of oil sand sample (38 mm in diameter) showing significant expansion (dark zones are looser).**

The proportion of each component of the specimen was determined. In a first step, the specimens were dried at laboratory temperature in a dessiccator with silica gel. The mass changes with respect to time were followed during drying. They became negligible after 4 months. The gravimetric water content was then determined with small values in the range 0.7-0.8 % in accordance with previous data of Dusseault (1980).

Secondly, the proportion of bitumen was determined by washing the sample in a Soxlhet cell at 250 °C with a solution consisting of 70 % toluene and 30 % methanol. After 2 weeks, the specimen became completely clean with all the bitumen dissolved It was dried into an oven at 60°C during several days. The proportion of bitumen was also determined after extraction by $CH_2Cl_2$. The sample masses at natural and dry states provided with both methods a proportion of fluid mass ($M_f/M$ where $M_f$ is the mass of bitumen and water and $M$ is the mass of specimen) in the range 14.8–15.2 % giving a gravimetric bitumen content ($M_b/M$ where $M_b$ is the mass of bitumen) of about 14-14.4 % with a water content $w$ ($w = M_w/M$ where $M_w$ is the mass of water) of about 0.7-0.8 %. Under the hypothesis of no gas and adopting $\rho_s = 2.65$ Mg/m$^3$ for the solid density, these values provide a porosity of 31.75 % (compared to the 31-35 % range derived from the combined use of in-situ density and neutron log data), a degree of saturation in bitumen of 94.7 % and in water of 5.3 %. Note that calculations carried out based on the determination of the volume of the specimen submitted to gas expansion provided degrees of saturation in bitumen $S_b = 71$ % and in water $S_w = 4$ %, corresponding to a degree of saturation in gas $S_g = 25$ % with a total porosity of 38 %. This indicated that the sample has been affected by a volume expansion of 4.6 – 10.1 % depending on the in-situ porosity. With a core porosity of 40 %, the disturbance index $I_D$ (Dusseault and van Domselaar 1982) lies in the range 8.57 – 22.6 % which indicates intermediate disturbance.

## X Ray Microtomography observation

A 5 mm diameter cylindrical oil sand specimen was investigated by using a X-Ray MicroTomography scanner with a spatial resolution of 6 μm. The workflow for image treatment and analysis of the reconstructed 3D volumes consisted of (1) visualizing, isolating and quantifying the resolved pore space and different phases, (2) estimating the grain size distribution. The volume was visualized and analyzed using the Avizo software package (version 6.2, VSG, France) and ImageJ software 1.44.

A 2D slice extracted of a 3D volume of 3 × 3 × 3 mm is presented in Figure 3. Sand grains (unit mass 2.65 Mg/m$^3$) appear in light grey, grey voids correspond to water and bitumen (with unit masses equal to 0.998 Mg/m$^3$ and 1.002 Mg/m$^3$, respectively). Black voids that correspond to a less dense medium are filled by gas. The image shows some dense areas composed of highly angular grains (light grey) and fluids (grey) that are separated by larger black pores (full of gas). Dense areas are characterised by an interlocked fabric made up grains with long and concavo-convex inter-grains contacts (Dusseault and Morgenstern 1978). The photos are comparable to that obtained by Leung et al. (1995) and Wong (2003) from back scattered SEM images.

The 3D microtomography (μCT) reconstruction of a cubic specimen of 3 × 3 × 3 mm side is presented in Figure 4 (top left). The Figure clearly evidences a crack of about 200 μm thickness (comparable to the average grain size) and about 1500 μm long on the horizontal plane. The other pictures show the packing of the denser zones (bottom left and right) and the black pores full of air (top right). It seems that some spherical small voids can be observed within the bitumen between the grains (bottom left and right). The inter-grains pores that are full of gas can be observed here in more details. They are limited by rather clean grains, with some bitumen remaining at inter-grains contacts.

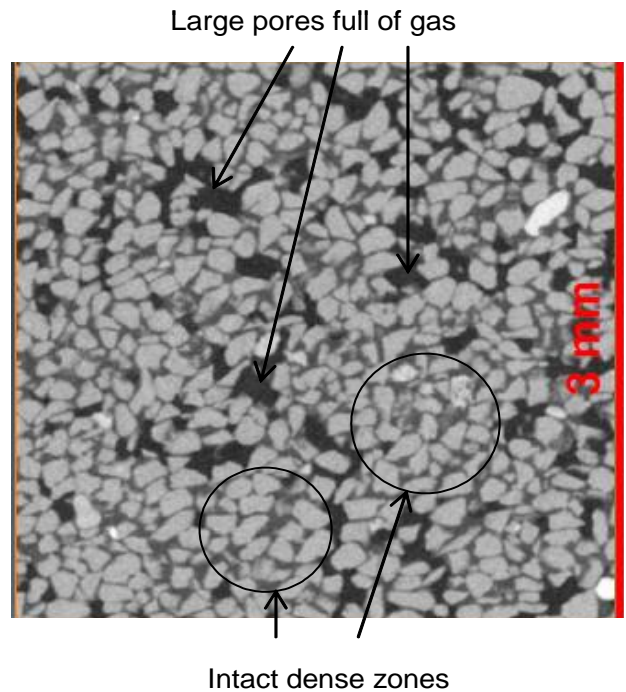

Figure 3. 2D X-Ray MicroTomography data of an oil sand specimen (3 × 3 mm side).

3D Image analysis was undertaken to extract quantitative values from µCT images. Image segmentation is a crucial step in image analysis. It aims to separate the different phases present in the raw image by assigning to each voxel the corresponding phase depending on its grey level. To do so, the images were previously filtered by assigning to each voxel the average grey value of its $5^3$ neighbours. This type of mean filter was found to be a good compromise between the elimination of noise and the smoothing effect. It results in an enhanced contrast and a better separation of the grey level peaks in the histograms. By applying thresholds th1 and th2 at the interpeaks values, a new composite image of the 3D phase distribution was obtained (Figure 4).

The different phase fractions extracted from the composite image can be expressed as $F_p = N_p/N_{img}$ and $F_f = N_f/N_{img}$, where $F_p$, $F_f$, $N_p$, $N_f$ and $N_{img}$ are respectively the resolved porosity fraction, the bitumen phase fraction, the number of voxels of the resolved porosity, the number of voxels of the bitumen phase and the total number of voxels in the image.

The computed proportions of the porosity full of fluid ($\phi_f = V_f/V = 30$ % where $V_f$ and $V$ are the volume of fluid in the pore and the volume of specimen respectively) and of gas ($\phi_g = V_g/V = 10$ % where $V_g$ is the volume of gas in the pore) give a total porosity of 40 %. Similarly, 3D image analysis carried out in the denser zones with no gas (observed in Figure 3 and Figure 4) provided a porosity value of 32 %. Interestingly, these values are in good agreement with that previously calculated from macroscopic measurements. This indicates that most of the pores full of air observed in µCT are due to gas expansion (there could be some voids corresponding to some small amount of gas naturally in the samples). The intact state of the oil sand is that observed in the dense zones indicated in both Figures, with a porosity value close to 31 %, the lower bound of in-situ measured porosity (31 – 35 %).

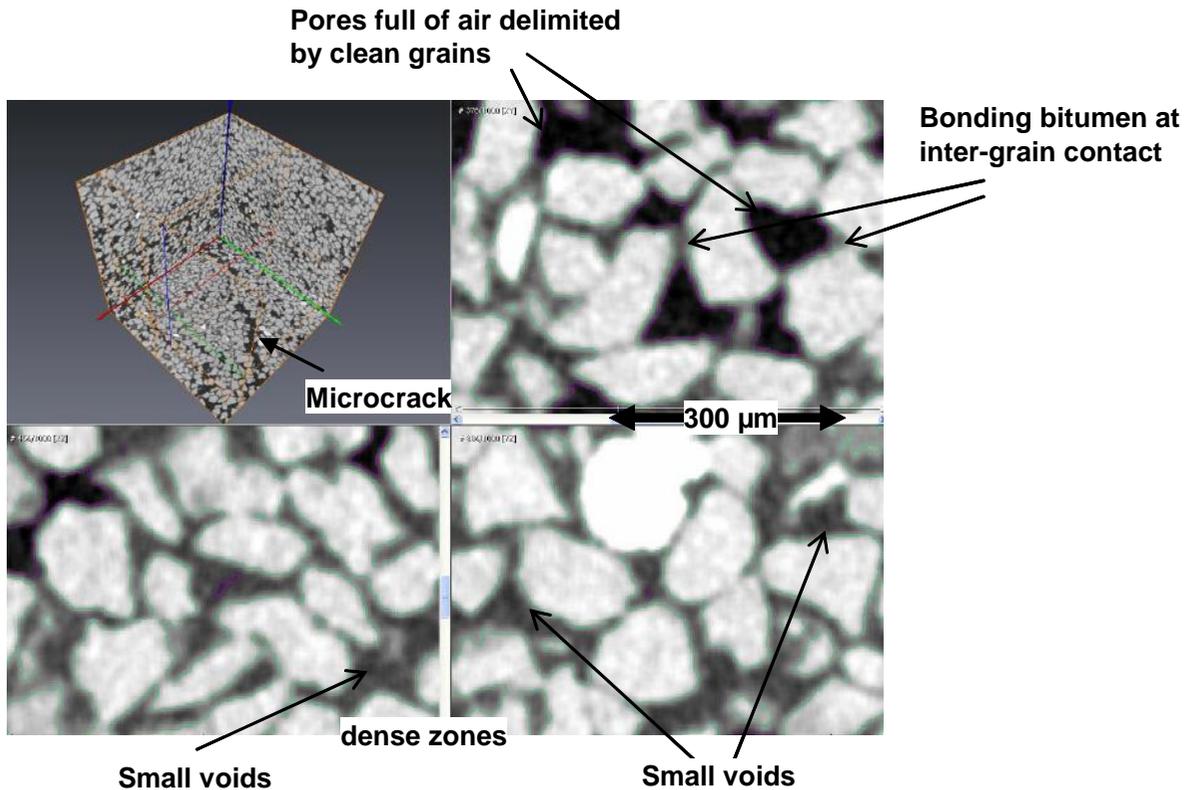

**Figure 4. 3D reconstruction from MicroTomography pictures (3D image: 3 x 3 x 3 mm side).**

3D Image analysis was also used to determine the grain size distribution of the grains observed in µCT. This was done by applying morphological operation (erosion and dilatation) on the solid phase. The curve obtained is compared in Figure 5 to the curves presented in Figure 1. The good correspondence observed indicates that the small volume investigated ($3 \times 3 \times 3$ mm) by µCT is fairly representative of the sand at macroscopic scale. The good grading of the curve around an average size relatively small (160 µm) is favourable in this regard.

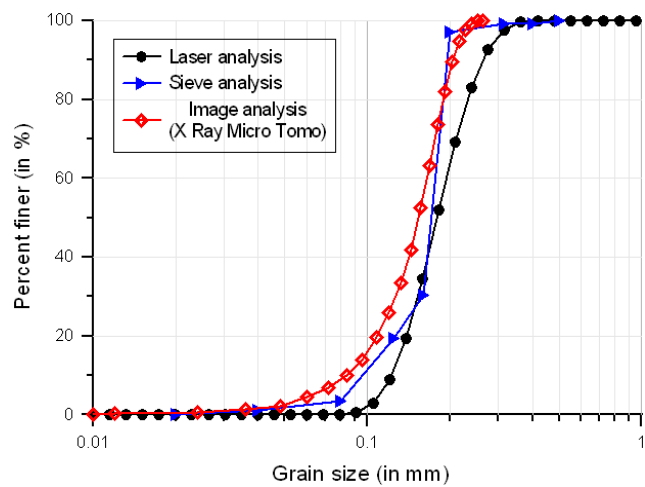

**Figure 5. Grain size distribution curve obtained from microtomography and image analysis**

## CryoSEM observation

In Cryo Scanning electron microscopy (CryoSEM), samples were frozen at -200°C into slush nitrogen prior to be introduced into a specific microscope airlock cooled at -170°C. The frozen samples were afterwards fractured and chromium metallised in the cooled airlock prior to be introduced into the observation chamber maintained at a temperature of -180°C. In this work, oil sand samples were observed by using a cryo-electron microscopy (ZEISS/SUPRA40-FEG module OXFORD CT1500) using the retrodiffused electron mode.

Quick freezing ensures very low water volume changes during freezing and satisfactorily preserves the initial microstructure of water saturated soils and rocks (Gillott 1973; Tovey and Wong 1973; Delage and Pellerin 1984). When fracturing frozen samples, the frozen fluid is solid and acts like an impregnation resin that maintains the particle in their initial arrangement. The fracture plane does not follow any weakness plane of the sample. Fractures in frozen materials are fragile and cross the various structure levels (Delage et al. 1982). This makes a big difference with the ductile fracture that would occur along the weakness planes of the microstructure within the bitumen when fracturing the sample under laboratory temperature.

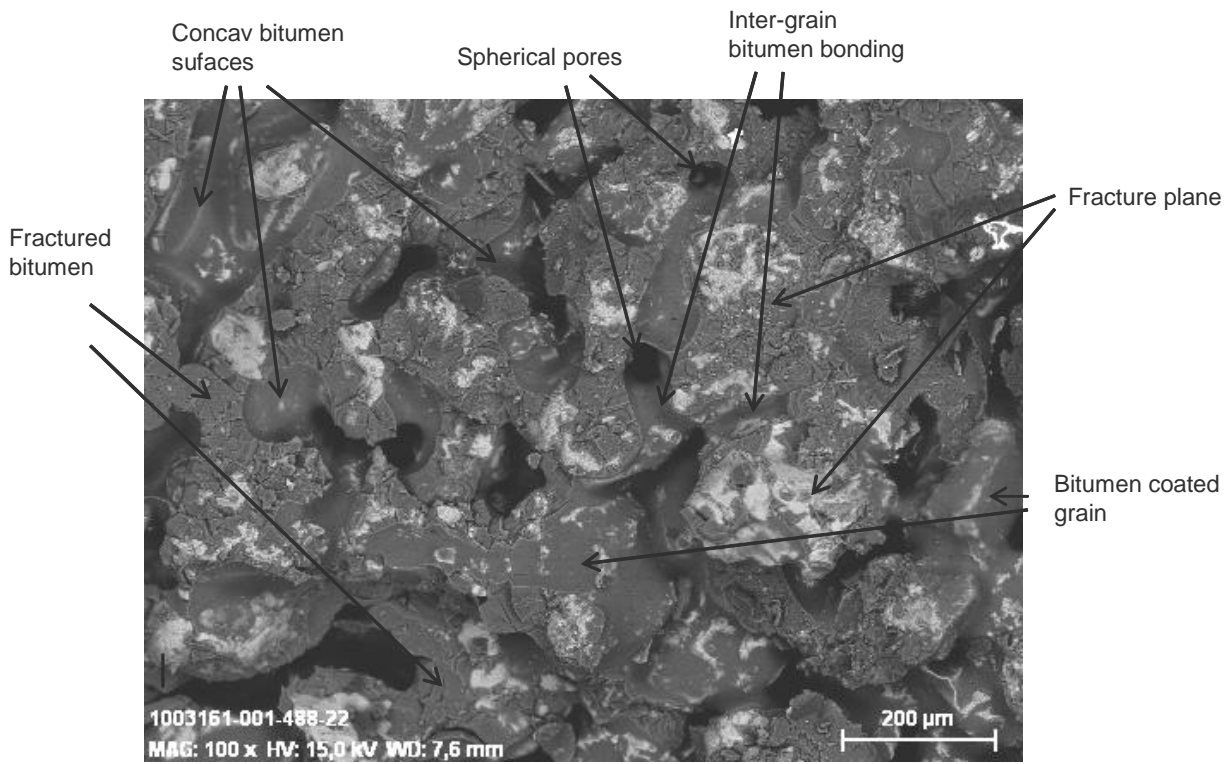

**Figure 6. CryoSEM observation of the oil sand sample. The fracture plane passes either along the grain edges or through the bitumen. Bitumen appears as a bonding agent connecting the grains together.**

Cryo-SEM observations were carried out in the denser intact areas evidenced above in μCT. The good quality of the freeze fractured planes is confirmed in the photo of Figure 6 in which one observes that the fracture plane passes either along the grain edges or through the bitumen. Sand grains along the fracture plane remain partially coated with bitumen chips that are remains from frozen bitumen that was directly linked to the grains before being torn off by freeze fracturing. The adherence between bitumen and sand grains is also attested by the observation of various bitumen coated grains. When not crossed by the fracture plane, bitumen presents nice and regular concave shapes with some appearance of menisci, compatible with a hydrophobic nature of the grain/bitumen interaction. Indeed, as suggested by Schmitt (2005) and also observed in µCT above, bitumen appears as a bonding agent connecting the grains together. This of course holds as far as bitumen remains solid enough, at low temperatures. Note also in Figure 6 some spherical voids (diameter between 50 and 80 µm) between the grains.

In the photo of Figure 7, the bitumen along the fracture plane sometimes appears to be cracked, a possible consequence of slight shrinkage during freezing. The rough aspect of grain surfaces that are characterised by many small holes (a well known feature in oil sands) is also observed in more details (bottom left photo) together with many chips of frozen bitumen that remained stuck along the grain surface during freeze fracturing (bottom right photo).

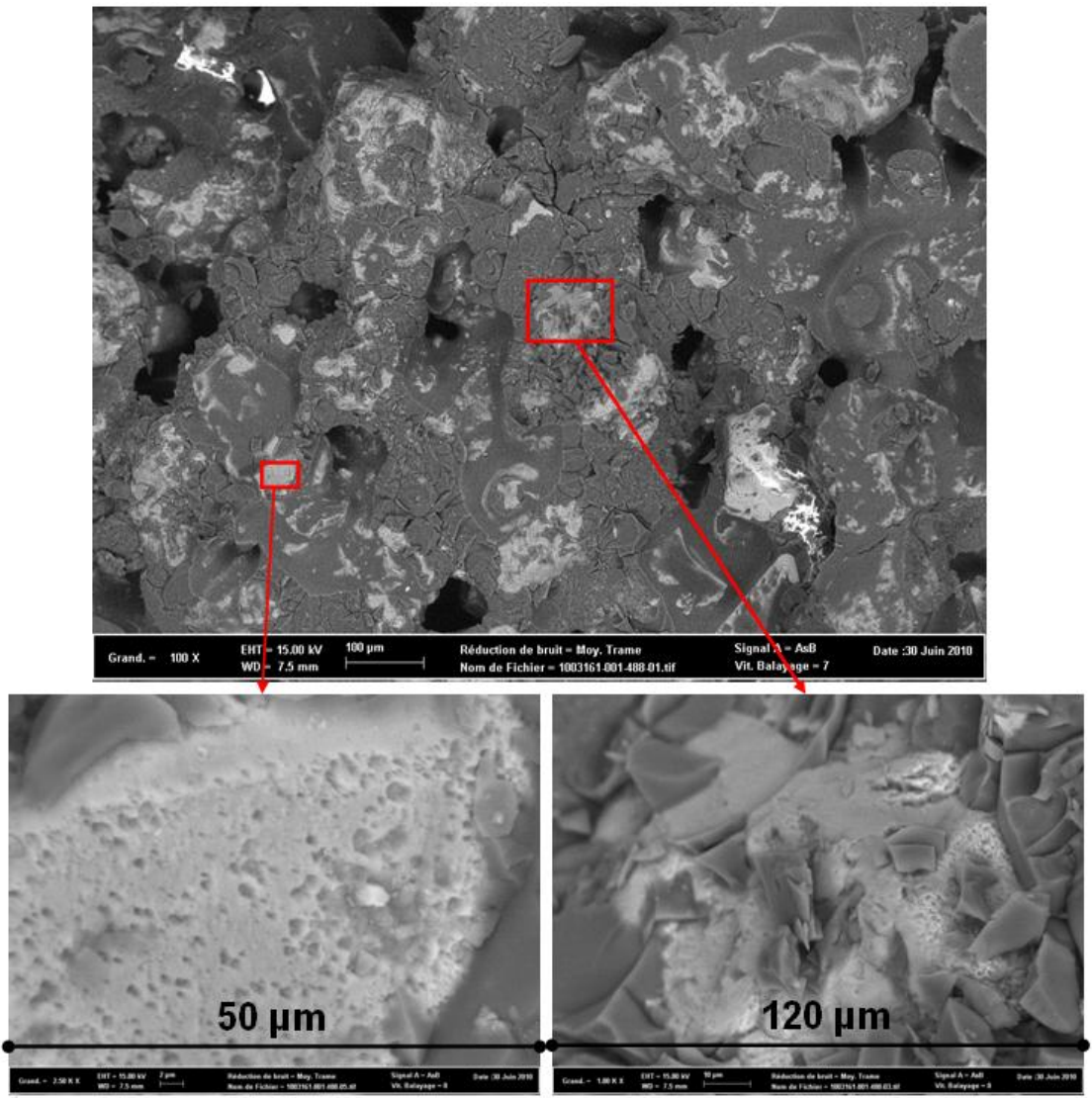

**Figure 7. Picture from the cryo-electron microscopy of oil sand sample. Bitumen along the fracture plane is divided into small pieces. On observes the rough aspect of grain surfaces with many small holes (bottom left photo) and bitumen chips (bottom right photo).**

In both photos, pores have regular forms delimited by either regular smooth concave surface of bitumen or by the grains themselves. Bitumen is either bonding or coating the sand grains.

At 75 m depth, the pores of oil sand are saturated with water. The pores observed in the bitumen matrix in Figure 6 and Figure 7 are part of a porous network that is responsible for the low water permeability of oil sands, estimated between $10^{-10}$ and $10^{-12}$ m/s by Dusseault (1980). This value is much less than that of sands (around $10^{-6}$ m/s) and is comparable to that of clays.

Punctual Energy-Dispersive X-ray Spectroscopy (EDS) analyses conducted on a grain and on bitumen (Figure 8) provided some information about silica (peaks of Si and O) and bitumen (peaks of carbon and sulphur).

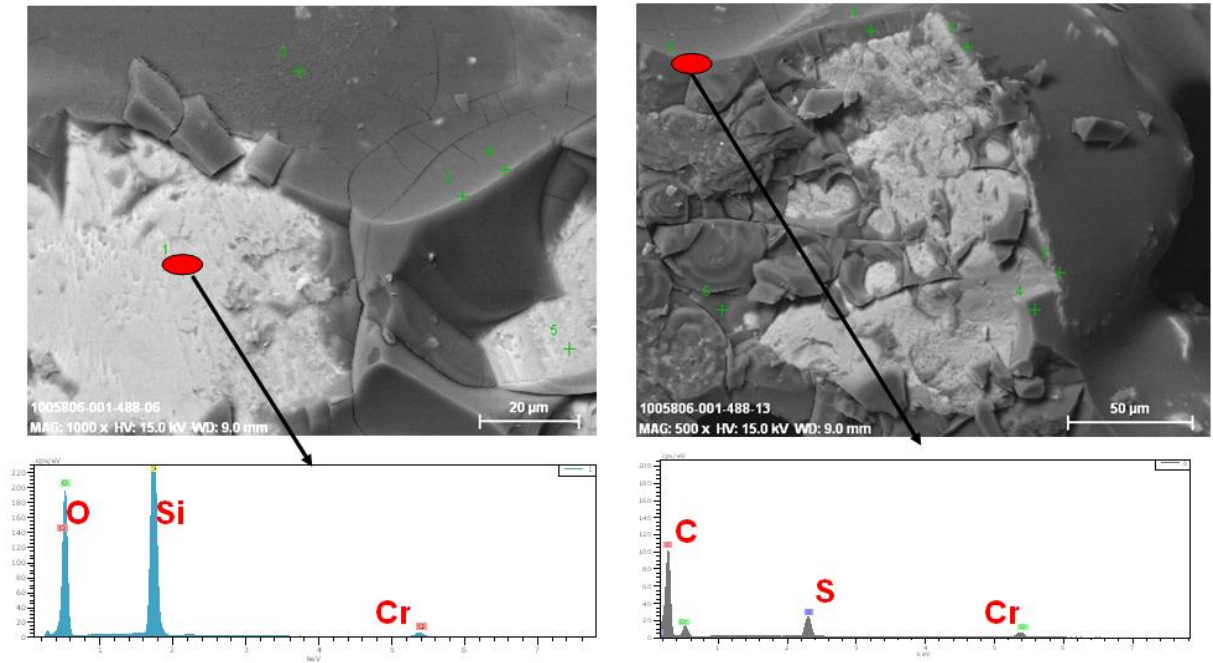

**Figure 8. EDS analysis on sand grain surface (left) showing the silica components (Si and O) and on liquid area (right) showing the bitumen components (C and S).**

Figure 9 presents the maps characterising the presence of carbon C, sulphur S and oxygen O in the photo presented in Figure 6. The C and S maps illustrate the presence of bitumen and bitumen coating whereas the O map evidences the presence of oxygen atoms in the silica grains that appear significantly clearer when not coated by bitumen.

As seen in Figure 10, some clay particles coating the grains that appear in light grey, with a granular shape, have been identified thanks to the determination of the local presence of Al, Si, O, K, Na and Mg. The clay is coating a grain with an angular shape and is apparently covered by bitumen coating. The presence of clay in the oil sand microstructure has been described in the model of Dusseault and Morgenstern (1978).

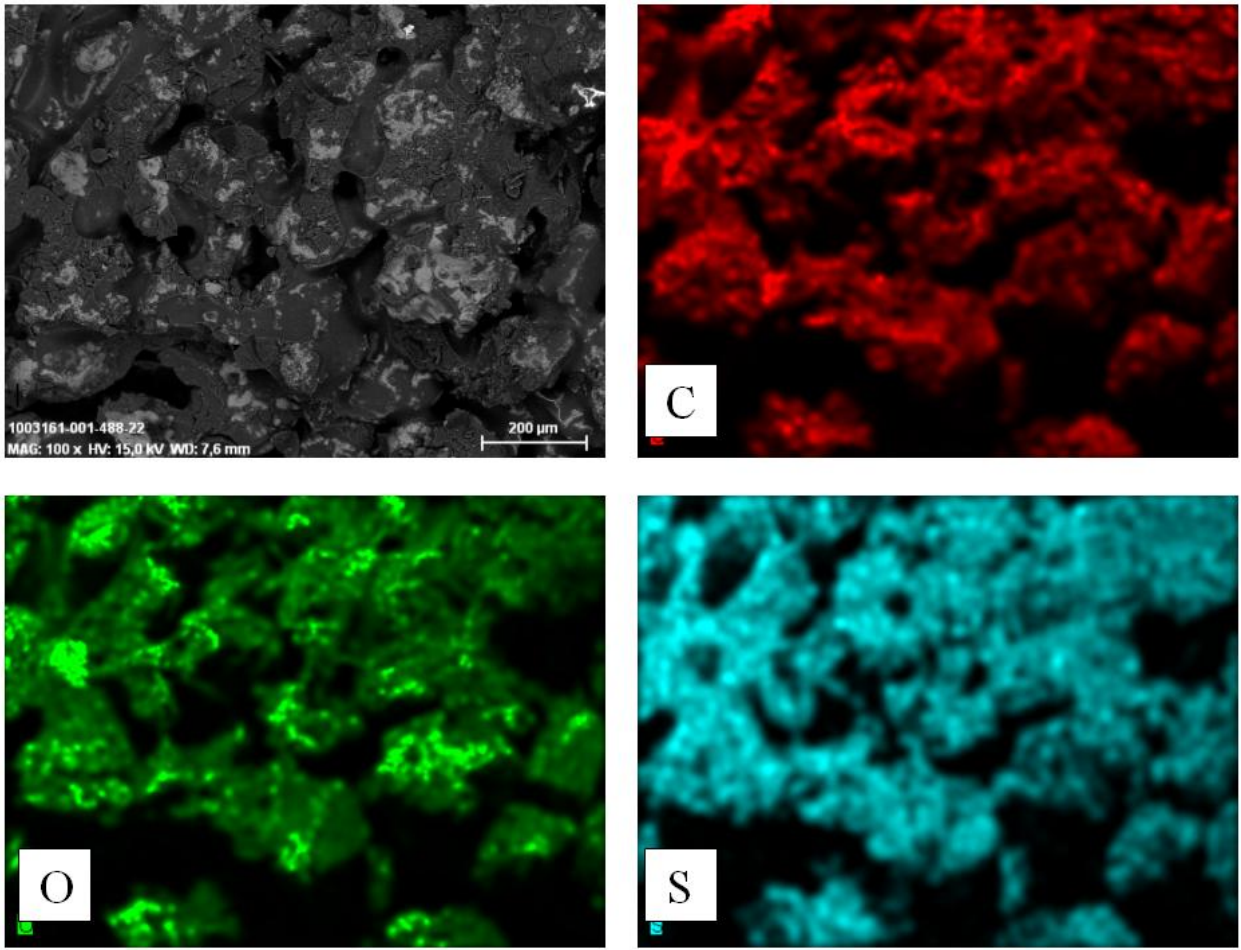

**Figure 9. EDS mapping of Carbon, Sulphur and Oxygen**

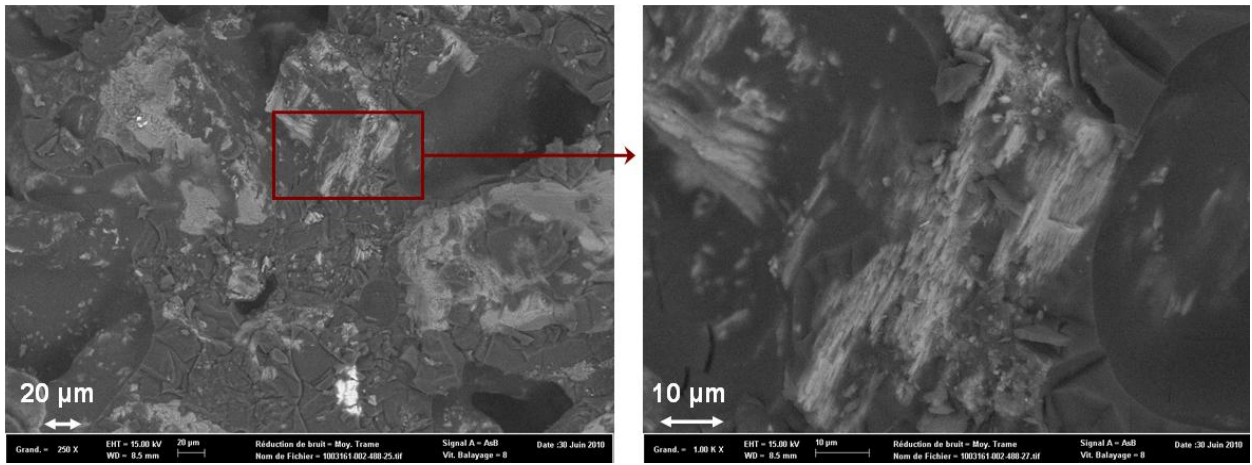

**Figure 10. Presence of clay in the oil sand sample.**

## Conclusions

The microstructure of oil sand samples from the McMurray formation in Athabasca was investigated by using X-Ray microtomography and Cryo-Scanning electron microscopy. 2D and 3D X-Ray µCT investigations provided a detailed description of the assemblage of the sand grains and of the presence of fluids and gas in the porosity. Artificial pores resulting from gas expansion were clearly identified given the significant difference in density between gas and liquids (bitumen and water). Image analysis carried out on digitalised µCT images evidenced that the porosity in zones with no gas was close to 32 %, a value also estimated from global measurements and in the range of the in-situ porosity measurements (31 - 35 %). The total porosity of the disturbed core sample was close to 38 % (from global measurements) or to 40 % (from µCT image analysis), showing that sample disturbance was characterised by a gas expansion of up to 10 % gas porosity and about 4.6 - 10.1 % total volume expansion.

CryoSEM observation on freeze fractured samples evidenced a regular and well organised microstructure in intact zones. The filling of the inter-grains porosity by bitumen has been observed in more details. Observation of bitumen coating on grains also suggested that the common hypothesis of a water sheath enveloping the grains was not confirmed in the Athabasca oil sand sample observed here. This observation is in accordance with the conclusions drawn by Zajic et al. (1981) from TEM observation on freeze fractured samples. The bitumen chips remaining on the grains along the freeze-fractured plane suggested that bitumen was strongly linked to the grains prior to be torn off by the fracture. The concave shape of the bitumen meniscii observed between the grains also suggested an oil-wet behaviour of the sample. Another important conclusion drawn from CryoSEM observations is the presence of nicely shaped rounded pores within the bitumen among the grains. This pore network, full of water under in-situ conditions, controls the macroscopic water permeability of the oil sand, a most crucial parameter in terms of steam assisted gravity drainage.

## Acknowledgements

The authors would like to thank Mrs. A.S. Gay and H. Marmet (IFPEN Solaize) for the CryoSEM picture and their recommendations. We also wish to thank Mrs. E. Rosenberg, M. M. Fleury from IFPEN Rueil and M. M. Mainguy from Total for many useful discussions.

## References

Butler, R.M. 1997. Thermal Recovery of oil and bitumen. GravDrain Inc., Alberta, Canada.
Clark, K.A. 1944. Hot water separation of Alberta bituminous sand. Transactions of the Canadian Institute of Mining and Metallurgy, **47**: 257–274.
Czarnecki, J., Radoe, B., Schramm, L.L. and Slavchev, R. 2005. On the nature of Athabasca Oil Sands. Advances in Colloid and Interface Science, **114–115**: 53–60.
Delage, P. and Pellerin, F.M. 1984. Influence de la lyophilisation sur la structure d'une argile sensible du Québec. Clay Minerals, **19**: 151-160.
Delage, P., Tessier, D. and Marcel-Audiguier, M. 1982. Use of the Cryoscan apparatus for observation of freeze-fractured planes of a sensitive Quebec clay in scanning electron microscopy. Canadian Geotechnical Journal, **19**(1): 111-114.
Dusseault, M.B. 1980. Sample disturbance in Athabasca oil sand. Journal of Canadian Petroleum Technology, **19**(4): 85–92.

Dusseault, M.B. and Morgenstern, N.R. 1978. Shear Strength of Athabasca Oil Sands. Canadian Geotechnical Journal, **15**: 216-238.
Dusseault, M.B. and van Domselaar, H.R. 1982. Unconsolidated sand sampling in Canadian and Venezuelan waters. Revista Technica Intevep, **2**(2): 165-174.
Gillott, J.E. 1973. Methods of sample preparation for microstructurel analysis of soil. In Soil Microscopy, Proceedings of the 4$^{th}$ International Working-Meeting on Soil Micromorphology, ed. G.K Rutherford, Kingston, pp. 143-164.
Hein, F.J. and Marsh, R.A. 2008. Regional geologic framework, depositional models and resource estimates of the oil sands of Alberta, Canada. In Proceedings of the World Heavy Oil Congress, Edmonton, Canada, 10 – 12 March, Paper 2008-320.
Leong S.K., Kry P.R. and Wong R.C.K. 1995. Vizualisation of deformation in unconsolidated Athabasca oil sand. Proc. SPE International Heavy Oil Symposium, 637-645, Calgary.
Mathieu, X. 2008. Géologie de l'Athabasca et de l'Orénoque. Technoscoop, **34**: 35-41.
National Energy Board (NEB). 2000. Canada's Oil Sands: A Supply and Market Outlook to 2015. Calgary, Alberta: October 2000.
Schmitt, D.R. 2005. Rock physics and time-lapse monitoring of heavy-oil reservoirs. In Proceedings of the International Thermal Operations and Heavy Oil Symposium, Calgary, Alberta, Canada, 1 - 3 November, Paper SPE 98075.
Shuhua, G. and Jialin, Q. 1997. Micro-structure model of some Chinese oil sand. Petroleum Science and Technology, **15**(9): 857-872.
Takamura, K., 1982. Microscopic structure of Athabasca oil sand. The Canadian Journal of Chemical Engineering, **60**: 538–545.
Tovey, N.K. and Wong, K.Y. 1973. The preparation of soils and other geological materials for the scanning electron microscope. In Proceedings of the International Symposium on Soil Structure, Gothenburg, Sweden, pp. 176-183.
Zajic, J.E., Cooper, D.G., Marshall, J.A. and Gerson, D.F. 1981. Microstructure of Athabasca bituminous sand by freeze-fracture preparation and transmission electron microscopy. Fuel, **60**(7): 619-623.
Wong, R.C.K. 2000. Shear deformation of locked sand in triaxial compression. ASTM Geotechnical Testing Journal, **23**(2): 158-170.
Wong, R.C.K. 2003. Strain-induced anisotropy in fabric and hydraulic parameters of oil sand in triaxial compression. Canadian Geotechnical Journal, **40**: 489-500.

**List of symbols**

$I_D$ : disturbance index

$\phi$ : in-situ porosity

$\phi_e$ : porosity measured in the laboratory

$M$ : mass of specimen

$M_f$ : mass of bitumen and water

$M_b$ : mass of bitumen

$M_w$ : mass of water

$w$ : water content

$\rho_s$ : solid density

$S_b$, $S_w$, $S_g$: degrees of saturation in bitumen, in water and in gas, respectively

$F_p$, $F_f$: resolved porosity fraction and bitumen phase fraction, respectively

$N_p$, $N_f$, $N_{img}$: number of voxels of the resolved porosity, of the bitumen phase and the total number of voxels in the image, respectively

$V_f$, $V_g$, $V$: volume of fluid in the pore, of gas in the pore and the volume of specimen respectively

$\phi_f$, $\phi_g$: computed proportions of the porosity full of fluid and of gas respectively